# Electronic states inside the gap of quasi-one-dimensional conductor NbS$_3$(I)

V.F. Nasretdinova[1] and S.V. Zaitsev-Zotov

*Institute of Radioengineering and Electronics of RAS, 125009 Mokhovaya 11, Moscow, Russia*

[1] +7(495)6293394, venera@cplire.ru

**Introduction**

NbS$_3$ is one of very few quasi-one-dimensional conductors that are already in the Peierls state at the room temperature [1]. NbS$_3$ (phase I) is also a crystal analog of widely studied polydiacetilene, organic polymer conductor demonstrating collective electron excitation - solitons [2].

Here we report our results of photoconduction spectra study in crystals of NbS$_3$(I) in a wide temperature range, electric field and additional illumination.

**Experimental methods**

The crystals were synthesized by direct reaction in vapor phase from components in a stoichiometric ratio with slight excess of sulphur. Typical sizes of the studied samples are 5 μm × 50 μm × 1 mm.

We use the standard lock-in amplifier technique for the photoconductivity measurements. The sample was mounted on a sample holder. The sample holder was placed inside an optical cryostat with quartz windows. A chamber for a sample was filled with the heat-exchange gas (nitrogen for T > 78 K or helium for T < 78 K) to reduce heating of the samples by light.

The measurements were done in the two-contact configuration in the voltage controlled mode. The noise level in our setup ~ $10^{-14}$ A at time constant τ = 1 s. A grid monochromator with a set of filters was used for spectral dependence measurements.

We studied spectra in a broad temperature range - from 4.2 K up to 250 K, at the electric field, $E$, applied to the samples, varying from 3-5 V/cm to $E ≈ 700$ V/cm. The upper temperature was limited by decrease of signal to noise ratio. We also undertaken measurements under additional illumination by LED. The parameters of the illumination were $ℏω = 1.16$ eV, $W/s$ ~ 0.1 mW/cm$^2$ where $ℏω$ is the energy of photons, $W/s$ is the light intensity per unit area.

**Results**

We observe a fast decrease of the photoconductivity with lowering the photon enery near $ℏω = 1$ eV that we ascribe to the enegy gap edge. This value of the gap is consistent with the previous measurements of transmittance [3].

We also observe three types of photoconduction spectral features at energies smaller the enegy gap value. The first one is the peak at $ℏω = 0.6$ eV. The peak is getting noticeable at temperatures below 100 K in the electric field $E ≥ 50$ V/cm. The additional illumination increases the amplitude of the peak and make it visible even at the liquid helium temperature range down to $T$ ~ 5 K (some heating by LED illumination may take place at the lowest temperatures). This peak is well reproducible.

The amplitude of this peak grows with until the saturation occurs when increasing applied voltage or the intensity of additional illumination by LED applied (see Figure 1). Peak is a well reproducible feature of all the measured samples at the described conditions.

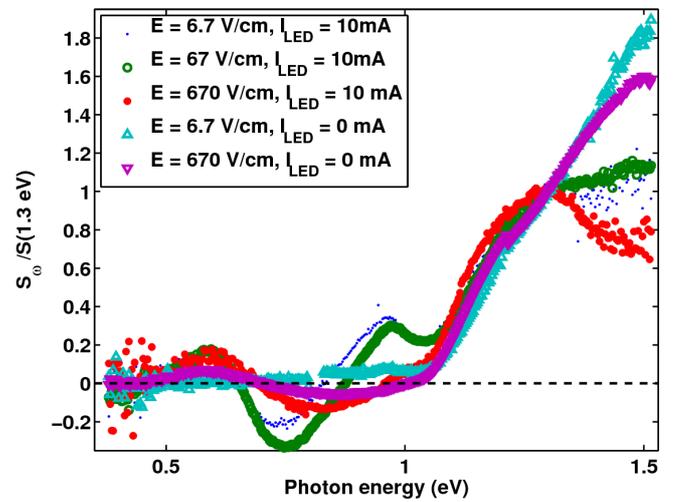

Figure 1. Photoconduction spectra of NbS$_3$(I) sample under different experimental conditions. The plotted quantity S$_ω$ is the measured photoconduction spectra δ$I$(ℏω)/$V$, where δ$I$(ℏω) is a photoconduction current, and V is the voltage applied to a sample, normalized by the number of incident photons $N = W(ℏω)/ℏω$ obtained from a bolometer spectra $W(ℏω)$ measured separately: $S_ω=δI(ℏω)/VN(ℏω)$.

The second feature we observed is the peak at ℏω = 0.9 eV. The amplitude of this peak vary from a sample to sampe (for example, at $T = 170$ K the amplitude of the feature varies by the factor 3). The peak amplitude non monotonously depends on the temperature, with a maximum at 210-220 K. Around $T = 100$ K the peak is small and produce a step-like feature in all the samples, and the peak disappears at much lower temperatures. The amplitude of the peak is suppressed by relatively low electric field $E$ ~ 100 V/cm, the characteristic voltage vary from a sample to sampe. Very similar effect was observed in o-TaS$_3$ earlier [4]. At $T = 78$ K increase of the additional illumination with LED leads to significant increase of the amplitude of the peak, whereas at $T = 200$ K the additional illumination does not affect noticeably the amplitude of the peak.

The third feature we found is the photoconductivity tail in the energy region ℏω = 0.7 – 1 eV. Small photoconductivity in the region between the peak at 0.6 eV and 0.9 eV is positive when the electric field applied to the sample is not large, of the order 5 V/cm (see Figure 1.). When a large electric field $E \sim 500$ V/cm is applied, value of the signal of photoconduction starts bleaching and photoconduction even becomes negative for some samples. When both the large electric field $E \sim 500$ V/cm, and intense additional illumination with LED are applied, the bleaching of photoconduction below zero for $0.7 < \hbar\omega < 1$ eV is observed in all the samples.

**Discussion**

The threshold-like dependence of the amplitude of the peak at the energy $\hbar\omega = 0.6$ eV on the electric field together with the increasing of the amplitude with LED illumination allows us suggest that this peak is concerned with polaronic or solitonic states that are expected to exist near the middle of the gap in the quasi-one-dimensional systems [5].

The electric-field dependent peak near the gap edge at the energy $\hbar\omega = 0.9$ eV resembles the exciton states like ones observed in polydiacetilene [6]. The difference is the opposite direction of their dependence on the electric field: in polydiacetilene the exciton states were observed to grow in the electric field, whereas they are decreasing in our measurements. Therefore we rule out this possibility. We suggest that this peak is produced by intrinsic defects of the structure, so-called stacking faults. Such defects are present in $NbS_3$ crystal structure [7], as well as in o-$TaS_3$, where a set of large peaks was also observed [4]. As the concentration of such defects is different in different samples, this could explain variation of the peaks amplitude among the samples.

The infrared bleaching of photoconduction at the energies $0.7 < \hbar\omega < 1$ eV corresponds to bleaching of the extra photoconduction produced by LED illumination and indicates that respective states inside the gap are responsible for relaxation of nonequilibrium quasiparticles. Further study is necessary to clarify details of this process.

*Acknowledgements*

This work was supported by the Russian Foundation for Basic Researches. The work was performed in the framework of the Associated European Laboratory "Physical Properties of Coherent Electronic States in Condensed Matter" of Institut Neel, Centre National de la Recherche Scientifique and Institute of Radio Engineering and Electronics, Russian Academy of Sciences.